\documentclass[]{aa}
\usepackage{graphicx}
\usepackage{times}
\usepackage{natbib}
\usepackage{amsmath}
\usepackage[usenames,dvipsnames]{color}

\begin{document}

\title{The globular cluster system of NGC 1316\thanks{Based on observations taken at the European Southern Observatory, Cerro Paranal, Chile, under the programme 082.B-0680, 076.B-0154, 065.N-0166, 065.N-0459.}}
    
\subtitle{IV.~ Nature of the star cluster complex SH2}

\author{
T. Richtler     \inst{1} 
\and 
B. Husemann \inst{6,2}
\and 
M. Hilker \inst{2}
\and 
T.~H. Puzia \inst{3}
\and 
F. Bresolin \inst{4}
\and
M. G\'omez \inst{5}
}
\offprints{T. Richtler}
\institute{
Departamento de Astronom\'{\i}a,
Universidad de Concepci\'on,
Concepci\'on, Chile;
tom@astro-udec.cl
\and
European Southern Observatory,
Karl-Schwarzschild-Str.2,
85748 Garching,
Germany
\and
Instituto de Astrof\'{i}sica,
Pontificia Universidad Cat\'olica de Chile,
Av.~Vicu\~na Mackenna 4860,
7820436 Macul, Santiago, Chile
\and 
Institute for Astronomy,
University of Hawaii,
2680 Woodlawn Drive,
Honolulu,
HI 96822, USA
\and
Departamento de Ciencias F\'{i}sicas, 
Facultad de Ciencias Exactas, 
Universidad Andres Bello, 
Fern\'andez Concha 700, 
7591538 Las Condes, Chile
\and
Max-Planck-Institut f\"ur Astronomie, K\"onigstuhl 17,
D-69117 Heidelberg,
Germany
}

\date{Received  / Accepted }

\abstract
 {The light  of the merger remnant NGC 1316 (Fornax A) is dominated by old and intermediate-age stars. 
 The only sign of
   current  star formation in this big galaxy is  the  H{\sc ii} region  
 SH2, an isolated  star cluster complex with a ring-like morphology and an estimated age of 0.1 Gyr  at a galactocentric distance of
   about 35 kpc.  A nearby  intermediate-age globular cluster, surrounded by weak line emission and a few more young star clusters,  is kinematically associated.
   The origin of this complex is enigmatic.} 
{We want  to investigate the nature of  this star cluster complex. The nebular emission lines permit  a metallicity determination which can discriminate between a dwarf galaxy or 
other possible precursors.}
{We used the Integrated Field Unit  (IFU) of the  VIMOS instrument at the Very Large Telescope of the European Southern Observatory in  high dispersion mode to study the morphology, kinematics, and
metallicity employing  line maps, velocity maps, and  line diagnostics of a few characteristic spectra.   }
{The line ratios of different spectra vary, indicating  highly structured H{\sc ii} regions, but define a locus of  uniform metallicity. The strong-line diagnostic diagrams 
and empirical calibrations  point to a nearly solar or even super-solar oxygen abundance. The velocity dispersion of the gas is
highest  in the region offset from the bright clusters. 
Star formation may be active on a low level.  There is evidence for a large-scale disk-like structure in the region of SH2, which would make  the 
similar radial velocity  
of the nearby globular cluster easier to  understand. }
{The high metallicity does not fit to a dwarf galaxy as progenitor. We favour the scenario of a free-floating gaseous
complex having its origin  in the merger 2 Gyr ago. Over a long period the densities increased secularly until finally the threshold for star formation was reached. 
SH2 illustrates how massive star clusters can form outside starbursts and without a considerable field population.  }

\keywords{Galaxies: individual: NGC\,1316 -- Galaxies: kinematics and dynamics -- Galaxies: star clusters}
\titlerunning{The nature of the star cluster complex SH2}

\maketitle

\section{Introduction}

NGC\,1316 (Fornax A)  in the outskirts of the Fornax galaxy cluster is prominent for its abundant signatures of previous galaxy interactions. Starting with the classical work of  \citet{schweizer80}, it has been intensively studied at various wavelengths from X-rays to the radio regime \citep{donofrio95,shaya96, mackie98, arnaboldi98, longhetti98, kuntschner00, horellou01, goudfrooij01b, goudfrooij01a, gomez01, kim03, bedregal06, nowak08,lanz10, goudfrooij12}.
 
Its globular cluster system has been investigated by \citet{richtler12a,richtler14} and the planetary nebulae population by \citet{mcneil12}. More photometric work on the cluster system has been presented by \citet{sesto16}. Although most of the  bright globular clusters show a narrow colour range, compatible with an age of about 2 Gyr, clusters as young as 0.5 Gyr have been found in NGC\,1316 \citep{richtler14}. However, a corresponding field stellar population of this age has not been identified until now. Even younger clusters are seen at only one place, the H{\sc ii} region SH2, detected by \citet{schweizer80} in the southern outskirts of NGC\,1316. \citet{richtler12b} presented the first  detailed description  based on HST/WFPC2 imaging in the F555W and F814W filters and VLT/FORS2 spectroscopy at 5\,\AA\ resolution.
 
This object is a star cluster complex containing about 100 young star clusters  within a volume of radius 0.35 kpc. A peculiar feature is another nearby ensemble of blue star clusters with some line emission, grouped around an intermediate-age massive globular cluster of NGC\,1316. A tentative interpretation of the nature of SH2 is that of an infalling gas-rich dwarf galaxy, as there are so many traces of previous infall processes through the rich shell system of the galaxy.   We are not aware of an obvious comparable object in any other galaxy (but see the remarks on the Ruby Ring in Sect.~\ref{sec:scenario}), which  does not mean that SH2 is too special to provide general insight into the process of star cluster formation.  
The purpose of the present contribution is to investigate in more detail the line emission, in particular the metallicity of the H{\sc ii}-gas, which is a key parameter for understanding the formation history of SH2 and which is  expected to be low in the case of a dwarf galaxy (e.g.~\citealt{smith12}).
 
We adopt the {supernova type-Ia distance of 17.8 Mpc quoted by \citet{stritzinger10}, corresponding to a distance modulus of $(m-M)_0=31.25$ mag}, but see also \citet{cantiello13} who quote a larger distance of 20.8 Mpc. One arcsecond corresponds to 86.3 pc at a distance of 17.8 Mpc.

\section{Observations and data reduction}
We observed the H{\sc ii} region SH2 roughly centred at RA(J2000): 03h 22m 38s and Dec(J2000): $-37^{o}$ 18\arcmin 47\arcsec\ on October 2012 in service mode with the integral-field unit (IFU) of the VIsible MultiObject Spectrograph \citep[VIMOS,][]{lefevre03} mounted on UT3 (Melipal) of the ESO Very Large Telescope in Chile. The high-resolution red (HRR) and blue (HRB) modes were used to cover the wavelength ranges 3900 -- 5335\AA\ at $R\!\simeq\!1700$ and 6315 -- 8600\AA\ at $R\!\simeq\!3100$, respectively. The field of view was chosen to be 27\arcsec$\times$27\arcsec\ with a sampling of 0.67\arcsec\  per pixel for both channels. Three individual exposures of 756 seconds  were taken for each channel and accompanied by three flat-field exposures and one arc-lamp exposure. A standard star for flux calibration was also observed  for each channel as part of the standard instrument calibration. The seeing during the observations was about 0.8\arcsec.

We performed all data reduction tasks with the independent \textsc{Py3D} data reduction package for fibre-fed IFU spectrographs  developed for the data reduction of the Calar Alto Large Integral Field Area (CALIFA) survey data \citep{husemann13}. It has been successfully applied to reduce VIMOS IFU data as well \citep[e.g.][]{husemann14} showing superior quality to the standard ESO pipeline.  With \textsc{Py3D} we first process the individual exposures performing bias subtraction, cosmic ray detection with \textsc{PyCosmic} \citep{husemann12}, fibre tracing and extraction, wavelength calibration, fibre-to-fibre transmission correction, and flux calibration. We also homogenise the highly varying spectral resolution across the VIMOS field to be 2.5\AA\ in both the HRB and HRR channel. Sky subtraction is done on each exposure by creating a high S/N mean sky spectrum from 200 fibres containing no significant object signal, which is subsequently subtracted from all fibres. Afterwards we combine the individually calibrated exposures to one datacube taking into account differential atmospheric refraction and bad pixels, such as cosmic ray hits or dead CCD pixels. 

The world-coordinate system of the data is set by assigning the HST coordinates of the bright star cluster 431-1 of SH2 \citep{richtler14} to the corresponding light weighted centre in the VIMOS data. 
For convenience we give all coordinates on the IFU in pixels, in
   the form X:Y for the centres of our spectral extractions (see
   Sect. \ref{sec:char} and Fig. \ref{fig:location}), which can be transformed as follows into right ascension (RA) and declination (DEC), both in degrees:
\begin{eqnarray*}
  \textrm{RA(J2000)} &=&  50.657833 - (16-y) \times 11.53\times 10^{-5}\\
\textrm{DEC(J2000)} &=& -37.313061 +(19 -x) \times 9.17\times 10^{-5}
\end{eqnarray*}

\section{Morphology and line maps}
\subsection{Appearance of the stellar cluster complex}
\citet{richtler12a} described the appearance of SH2 on the basis of ground-based and HST imaging. 
The brightest cluster (431-1) has an age of about 0.1 Gyr, based on its colour,  and an estimated mass of $1.9\times10^5 M_\odot$, if we adopt solar metallicity. The colours indicate a mix of populations with many clusters being even younger, but the extreme crowding prohibits precise measurements. A ring-like structure of the brightest clusters is discernible. At a distance of 12\arcsec\ towards the south-west a second assembly of clusters of about the same age exists, but this time the dominant and apparently central object is an intermediate-age globular cluster of NGC\,1316. \citet{richtler12a} estimated an age of $\sim\!2$ Gyr and a mass of about $10^6 M_\odot$. It cannot be excluded that this is a chance projection. 

\begin{figure*}[t]
\begin{center}
\includegraphics[width=\linewidth]{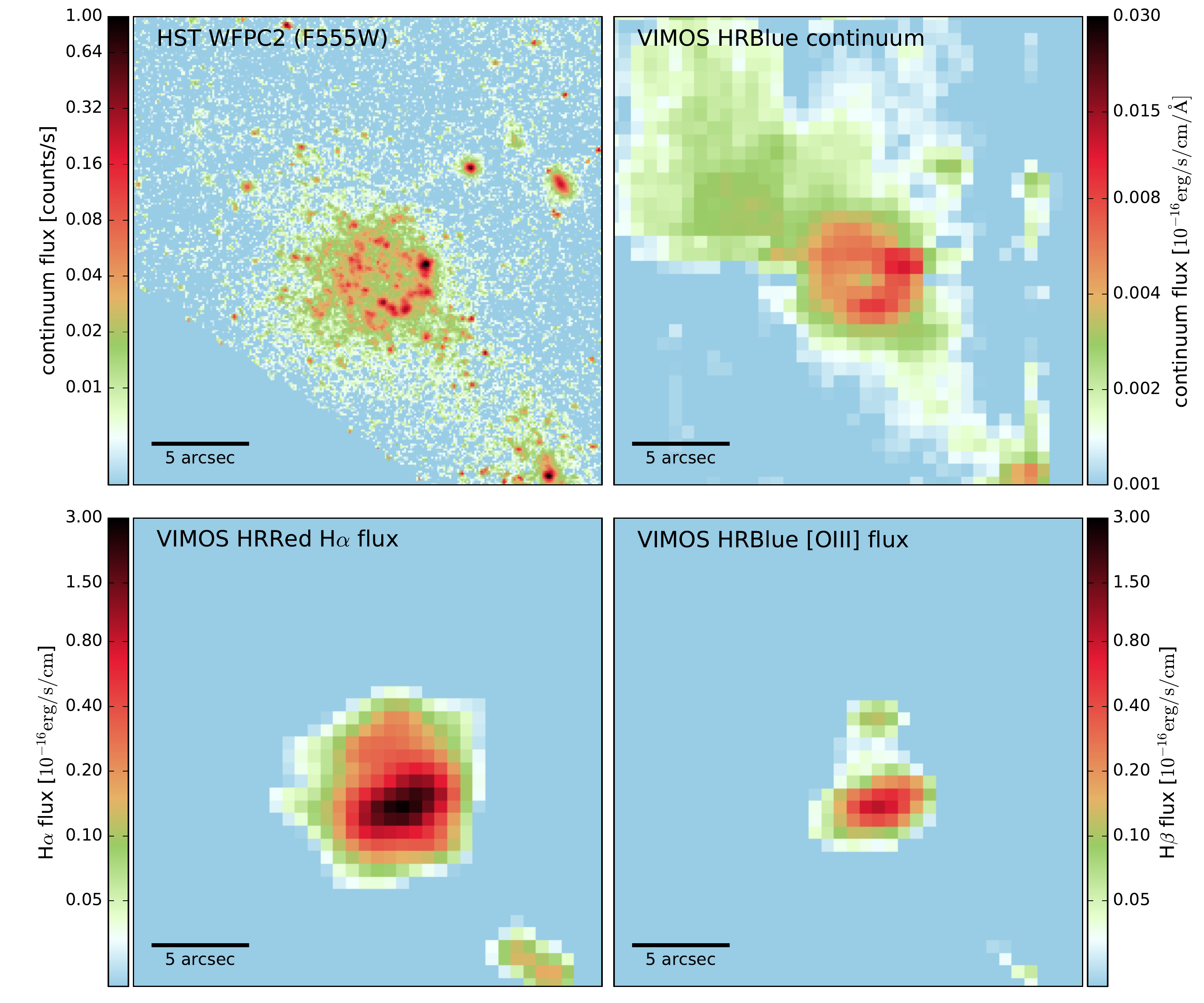}
\caption{HST image and line maps of SH2. All panels have a size of 27\arcsec$\times$27\arcsec\ corresponding to $2.33\times2.33$\,kpc$^{2}$. The 5\arcsec\  scale bar corresponds to 432\,pc. ({\it Upper left panel}): F555W HST-WPC2 image, matching the line and continuum maps.~({\it Upper right panel}): Continuum map from the HRB grism spectra, using median values. ({\it Lower left panel}): H$\alpha$ flux map from the HRR instrument set-up. ({\it Lower right panel}): Map of the [O{\sc iii}]-line at 5007\AA\  using the HRB set-up. In all frames north is towards the top and east is left. 
}
\label{fig:maps}
\end{center}
\end{figure*}

\subsection{Line maps}
A census of line emission regarding extent and intensity in the field of SH2 is provided by Fig.\ref{fig:maps}. The upper left panel shows an HST-image, where the structure of SH2 is clearly visible (PI: A. Sandage, GO-7504). The upper right panel shows a continuum map, using the median value of the data cube of a given pixel in the HRB mode. Unfortunately, the exposure times are not sufficient to spectroscopically explore the stellar continua. Some of the very faint structure in the upper left quadrant may be flat-field residuals, but the ring of star clusters of SH2 is clearly visible.   A faint population between SH2 and the globular cluster 429  is also visible towards the south-west. The colour map of \citet{richtler12b} indicates a somewhat redder colour than that of SH2, which is expected if a faint blue population is superimposed on a bright red galaxy background/foreground.   At the same time, it is a strong argument that the cluster assembly around 429 still belongs to the SH2 complex. 

The lower left panel is an H$\alpha$ map from the HRR set-up. The dynamical settings of the display  make even the faintest structures visible. In particular, it shows that there is no H$\alpha$ line detected above an S/N $>$3 between SH2 and the 429 region. The point of maximum H$\alpha$-intensity is a rather well-defined peak and coincides more or less with the most southern point of the ``ring''.

In the region of 429,   there is a double structure: one peak coincides with the location of 429 itself, the other peak is shifted by about 3\arcsec\ to the north-east. 
 A 2 Gyr old star cluster as a source of ionising radiation might appear surprising (see further details in Sect. \ref{sec:429}).

The lower right panel shows the morphology of the [O{\sc iii}]-5007\AA\ emission with the same absolute flux scaling and also clipping at a threshold of S/N $>$ 3.
 Its morphology closely resembles  that of the H$\alpha$ emission. The main part of the [O{\sc iii}] emission traces the southern part of the ring, peaking roughly at the same location as H$\alpha$. The presently ionising (invisible) young stars still seem to follow the overall distribution of the 0.1 Gyr clusters.

\section{Velocities and velocity maps}

The knowledge of radial velocities in this region so far stems from a few sources: \citet{schweizer80} quoted $\Delta v_{\rm rad}\simeq-101$ km/s as the offset to the systemic velocity of NGC\,1316, which is in excellent agreement with the H{\sc i}-velocity of SH2 from \citet{horellou01} of 1690 km/s. \citet{richtler14} measured radial velocities from medium dispersion VLT/FORS2 spectra for the objects 431 and 429 from the stellar continuum and emission lines. The results are for 429: $1709\pm12$ km/s for the emission lines and $1685\pm20$ km/s for the stellar continuum, where the quoted uncertainties are internal errors of the used code.~The velocity from emission lines for 431 is $1670\pm20$ km/s. 

\begin{figure*}[t]
\begin{center}
\includegraphics[width=\linewidth]{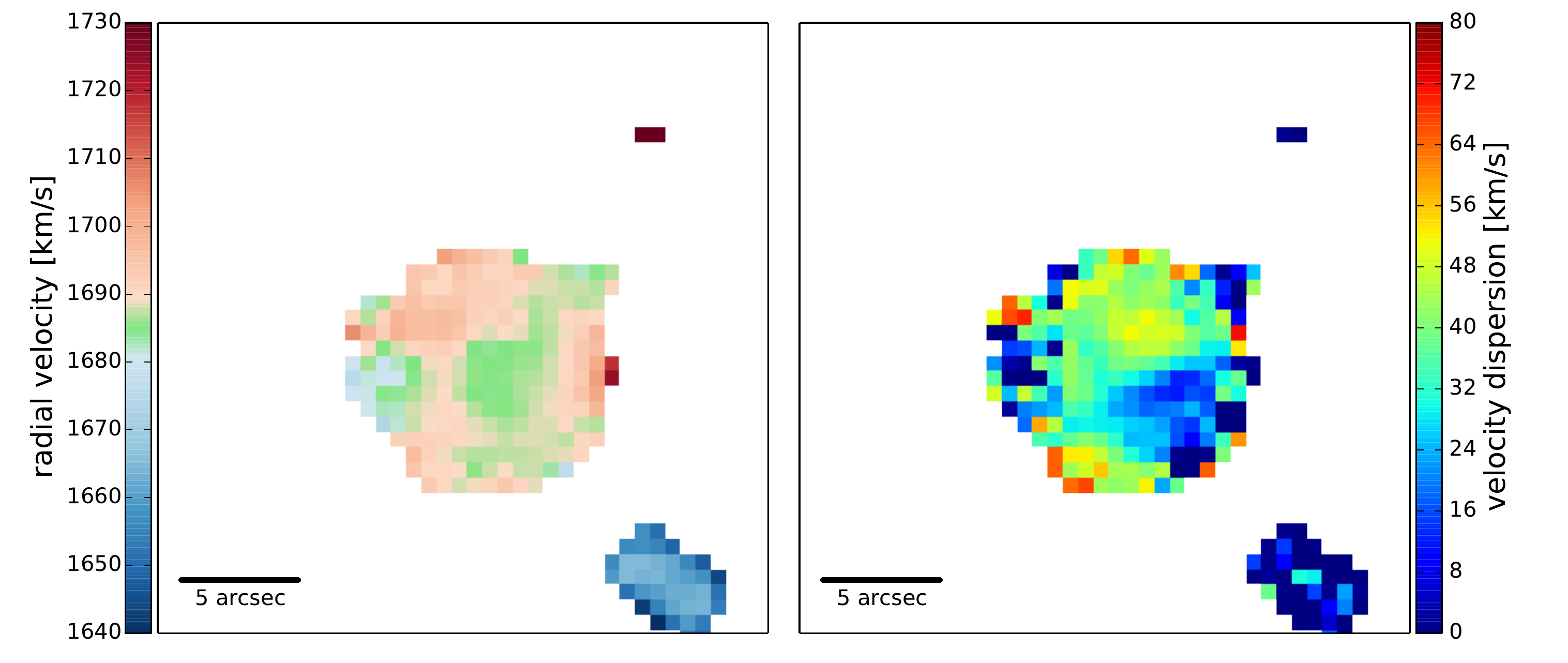}
\caption{({\it Left panel}): Radial velocity map of the SH2 region. Neither rotation nor velocity gradients are visible. The globular cluster 429 is offset by about $-30$ km/s from the main body of SH2. The object in the north-western region (also visible in Fig.\ref{fig:maps}) is a background galaxy without emission lines. The velocity is therefore
not valid and the redshift is unknown.  ({\it Right panel}): Velocity dispersion map of SH2. Except for the south-western region that is also the zone of the brightest H$\alpha$-emission, the dispersion is higher than expected from simple stellar-dynamical considerations. An interesting feature is that the spot with the highest dispersion coincides with the highest [O{\sc iii}]-flux intensity (see lower right panel of Fig.~\ref{fig:maps}).}
\label{fig:velocitymap}
\end{center}
\end{figure*}

In Fig.~\ref{fig:velocitymap} we show the complete kinematics maps, produced by the software {\it PyParadise} (\citealt{husemann16}, Husemann et al. in prep.), which is an improved version of {\it Paradise} \citep{walcher15}. The left panel of Fig.~\ref{fig:velocitymap} displays the maps of radial velocities, the right panel displays the velocity dispersion.~The pixels are not independent because of the seeing, which is slightly larger than the pixel size, but a velocity gradient or rotation is not visible. In the colour scaling of the radial velocity map, the green zone resembles approximately the interior of the ring-like arrangement of star clusters. The intermediate-age globular cluster 429 is offset by about $-30$ km/s with respect to the main body.  While the velocity for 431 is in good agreement with that from the FORS spectra, the agreement in the case of 429 is not as good.~It is, however, remarkable that 429, as a cluster from the NGC\,1316 system with a large velocity dispersion, does not show a larger velocity difference. This could be understood if the entire region were disk-like with only a small extension along the line of sight. 
 
The velocity dispersion map shows an interesting pattern. It is low in the region of the bright clusters and distinctly higher in the northern part. This spot is also roughly coincident with the secondary maximum of the [O{\sc iii}]-emission (see bottom right panel in Fig.~\ref{fig:maps}). Moreover, the line ratios of this region differ from those of the cluster region. We expect a  stellar projected velocity dispersion of the order of $\sigma^2_{\rm LOS} = \beta\, {G {\cal M}/R}$, where ${\cal M}$ is the mass, $R$ a characteristic radius, and $G$ the gravitational constant. The factor $\beta$ depends on the density slope of the tracer population, the orbit anisotropy, and the mass profile. Under isotropy, $1/\sqrt{3}$ is a good guess. The dominant mass component is the H{\sc i}-mass with about $10^7 M_{\odot}$ \citep{horellou01}, while the stellar mass has been estimated to be about $10^6 M_{\odot}$ \citep{richtler12b}. The concentration of the total mass is unknown and not resolved by VLA observations. Setting $R=50$ pc and ${\cal M}\!=\!10^7 M_{\odot}$, we expect a dispersion of about 17 km/s, which fits the southern part of the velocity map but not the high values that characterise the [O{\sc iii}]-emission. Because stellar dynamics cannot produce a dispersion as high as 40 km/s, gas dynamics might be a better explanation. Shocks from supernovae and stellar winds might have left their imprints on the interstellar gas. Our data are not appropriate to investigate these possibilities in more detail. Only data with higher S/N to investigate the stellar population itself in combination with higher spectral resolution would give more insight.

\section{Spectra and line fluxes}

\subsection{Fluxes and uncertainties}
We measure the line fluxes using the IRAF command {\it splot}, in particular the option of fitting and subtracting the underlying stellar spectrum.~The treatment of the stellar contribution (lines and continuum) is rather uncritical for the grism HRR because the blue stellar population contributes only  very little flux. For example, the H$\alpha$ flux is not notably affected, while the H$\beta$-fluxes measured in HRB are higher by about 2\% if the stellar spectrum is not subtracted. Because the fainter lines are more strongly affected (by about 5\%; this systematic difference disappears largely when line ratios are considered), we measured all fluxes after subtraction of the continua which have been modelled by third-degree polynomials. The fluxes and their uncertainties  are given in Tables~\ref{tab:fluxesHRB} and \ref{tab:fluxesHRR}.~The quoted uncertainties are estimated through repeated measurements and do not reflect uncertainties in the flux calibration.  For [O {\sc iii}] and [N {\sc ii}], only the brighter
line of the respective doublet is given. The lines  [O {\sc iii}] 4958\AA\ and [N {\sc ii}] 6548\AA\ are always fainter by a factor of 0.34.

 \begin{table*}
\centering
\caption{Line fluxes in several selected spectra obtained with HRB. The fluxes are in units of $10^{16}$ erg/s/cm$^2$. }
\label{tab:fluxesHRB}
\begin{tabular}{rccccc}
\hline
\hline
ID & spectrum  & H$\delta$ & H$\gamma$  & H$\beta$ & [O {\sc iii}]  \\
 wavelength &  & 4101  & 4340             & 4861         & 5007  \\
\hline
1 & 21:21  (sum)            & 2.44$\pm$0.04   &6.21$\pm$0.1    &16.50$\pm$0.2   &11.63$\pm$0.2 \\
2 &17:17    (average)           &   0.10$\pm$0.005  & 0.21$\pm$0.005  & 0.52$\pm$0.01 & 0.46$\pm$0.008  \\
 3& 4:5  (average)    &      ? & ?   & 0.037$\pm$0.002  &   0.026$\pm$0.001   \\
4 & 25:17 (average)  &         0.012$\pm$0.002        &   0.017$\pm$0.002        &      0.039$\pm$0.002                &  0.058$\pm$0.003        \\
5 & 22:15 (average)   &        0.013$\pm$0.002   &   0.025$\pm$0.001       &   0.074$\pm$0.003     &   0.026$\pm$0.002       \\
6 & 20:15 (average)  &       0.03$\pm$0.002  &  0.065$\pm$0.002            &    0.20$\pm$0.003     &   0.088$\pm$0.002            \\
7 & 21:20 (average)  &        0.01$\pm$0.002  &  0.03$\pm$0.002            &    0.077$\pm$0.003  &    0.027$\pm$0.002            \\
\hline
\hline
\end{tabular}
\end{table*}

\begin{table*}
\centering
\caption{Line fluxes in several selected spectra obtained with HRR.  The fluxes are in units of $10^{16}$ erg/s/cm$^2$. }
\label{tab:fluxesHRR}
\resizebox{18cm}{!}{
\begin{tabular}{rccccccc}
\hline
\hline
ID & spectrum     &H$\alpha$ &[N {\sc ii}] & He {\sc i} &    [S {\sc ii}] &  [S {\sc ii}] &  [Ar {\sc iii}]  \\
 wavelength  & & 6563      & 6582 & 6678 & 6715 & 6731 & 7135 \\
\hline
 1 & 21:21  (sum)                     & 60.23$\pm$0.3 & 21.18$\pm$0.2 & 0.52$\pm$0.03 &  8.77$\pm$0.1 & 6.03$\pm$0.1 &  1.16$\pm$0.05 \\
2 & 17:17    (average)           & 2.05$\pm$0.02 & 0.71$\pm$0.02 & 0.02$\pm$0.002 & 0.22$\pm$0.01 & 0.16$\pm$0.01 & 0.05$\pm$0.002 \\
3 & 4:5  (average)         & 0.082$\pm$0.004 & 0.028$\pm$0.002  & \dots & 0.006$\pm$0.001 & 0.007$\pm$0.001  & 0.002$\pm$0.0003 \\
4 &  25:17 (average)             &   0.087$\pm$0.002        &  0.033$\pm$0.002    &\dots     &     0.02$\pm$0.001       &     0.012$\pm$0.001  & \dots       \\
5 &  22:15 (average)                 &     0.14$\pm$0.002      &   0.052$\pm$0.002 & 0.005$\pm$0.0005       &    0.03$\pm$0.001        &     0.02$\pm$0.001    & 0.001$\pm$0.0002     \\
6 & 20:15 (average)        &   0.61$\pm$0.01   & 0.21$\pm$0.003     &    \dots & 0.08$\pm$0.002   & 0.06$\pm$0.002  &  0.01$\pm$0.001        \\
 7 & 21:20 (average)          &   0.23$\pm$0.01   & 0.08$\pm$0.003     &    \dots     & 0.048$\pm$0.002  &  0.03$\pm$0.002 & \dots       \\
\hline
\hline
\end{tabular}
 }
\end{table*}

\subsection{Extraction of some characteristic spectra}
\label{sec:char}
Although the substructure within SH2 is not very much larger than the seeing, there are clearly differences  in the line strength ratios.
To have a sample of  characteristic spectra which represents the morphology and the existing line strength ratios,  we chose the following  (locations are given
as x-y-coordinates): 
\begin{enumerate}
\item {\bf global}: extraction around 21:21 with a radius of 8 pixels. This sums over 197 spaxels;
\item {\bf spot of brightest H$\alpha$-emission}: extraction around 17:17 with a radius of 2 pixels. Average over 13 spaxels;
\item {\bf globular cluster 429}: extraction around 4:5 with a radius of 2 pixels. Average over 13 spaxels;
\item {\bf spot of the highest [O{\sc iii}]/H$\beta$ ratio}: extraction around 25:17 with a radius of 1 pixel. Average over 5 spaxels. This position corresponds
to the planetary nebula No. 397 in the catalogue of \citet{mcneil12};
\item {\bf spot of the lowest [O{\sc iii}]/H$\beta$ ratio}: extraction around 22:15 with a radius of 1 pixel. Average over 5 spaxels;
\item {\bf star cluster 431}: extraction around 20:15 with a radius of 1 pixel. Average over 5 spaxels; 
\item {\bf north-eastern region of low H$\alpha$-emission}: extraction around 21:20 with a radius of 2 pixels. Average over 13 spaxels. 
\end{enumerate}
 
\begin{figure}[t]
\begin{center}
\includegraphics[width=0.5\textwidth]{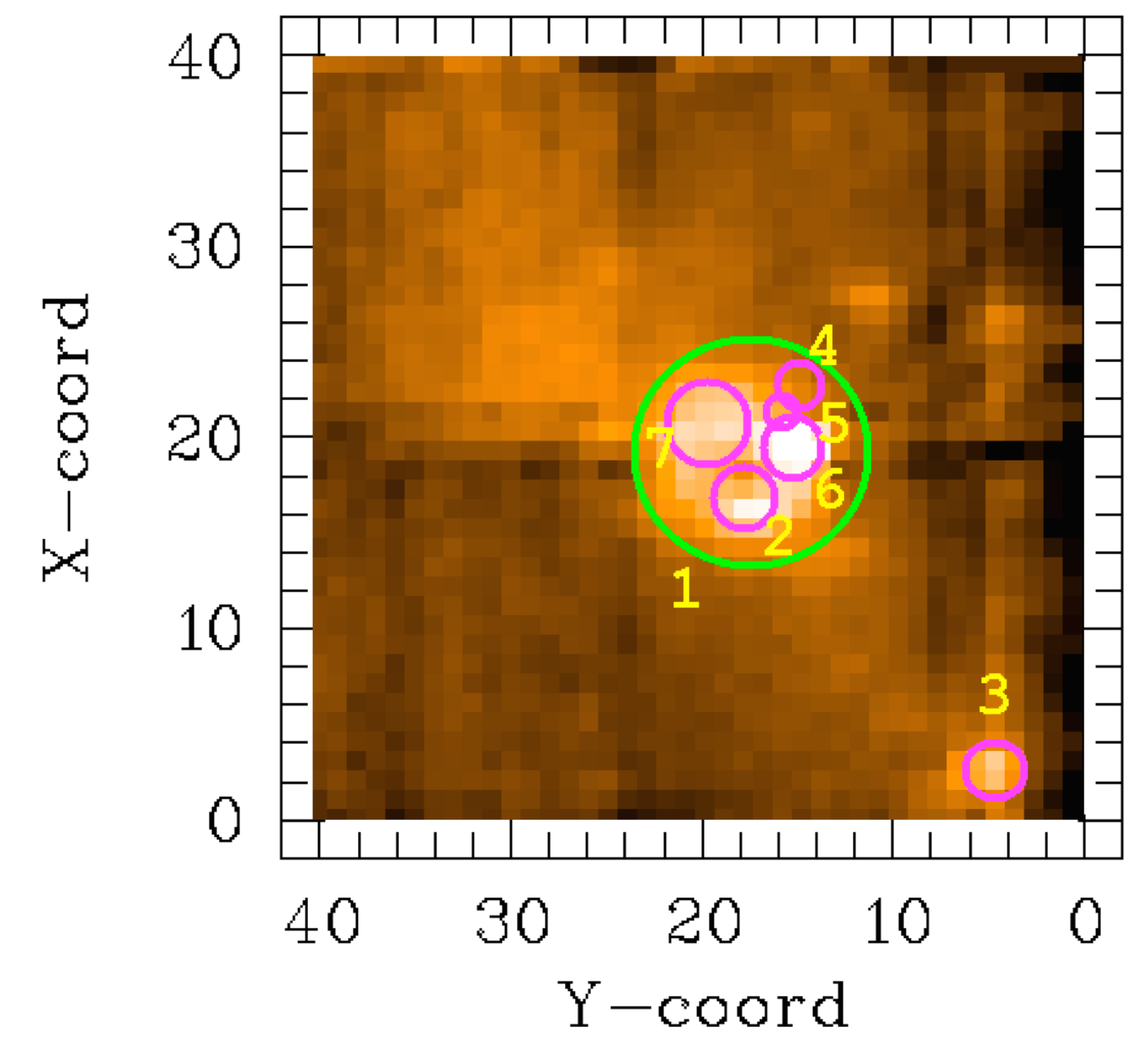}
\caption{The approximate locations and approximate sizes of the extraction radii of our characteristic spectra. The image is the median of the HRblue data cube. The unusual X-Y-configuration reflects the axes on the original data cube. }
\label{fig:location}
\end{center}
\end{figure}

Figure~\ref{fig:location} shows the locations of the spectra and the approximate sizes of the extraction radii. Figure \ref{fig:spectrum_total} displays the
total spectrum. We do not show the others because of their similar appearance.
  
\subsection{H$\alpha$-flux and star formation rate}
We measure the H$\alpha$-flux in spectrum 1 that averages over a radius of 8 pixels (see Section~\ref{sec:char}). The flux is $60.5\times10^{-16}$ erg/s/cm$^2$ and the luminosity $2.17\times10^{38}$ erg/s.~Adopting for the star formation rate the conversion of \citet{panuzzo03}, 
 SFR/H$\alpha$-luminosity = $7.05\times10^{-49} M_\odot/$erg, the SFR is $2.2\cdot10^{-4} M_\odot$/yr or $3\cdot10^{-4} M_\odot$/yr/kpc$^2$, much lower than a starburst, and still lower than ``normal'' star formation rates in normal galaxies
\citep{wuyts11}.
 Clearly, there must have been a starburst 0.1 Gyr ago with a much higher H$\alpha$-flux of which we now only see the
aftermath.

\section{Diagnostic graphs}

\subsection{Models from \citet{dopita13}}
\begin{figure*}[t]
\begin{center}
\includegraphics[width=0.8\linewidth]{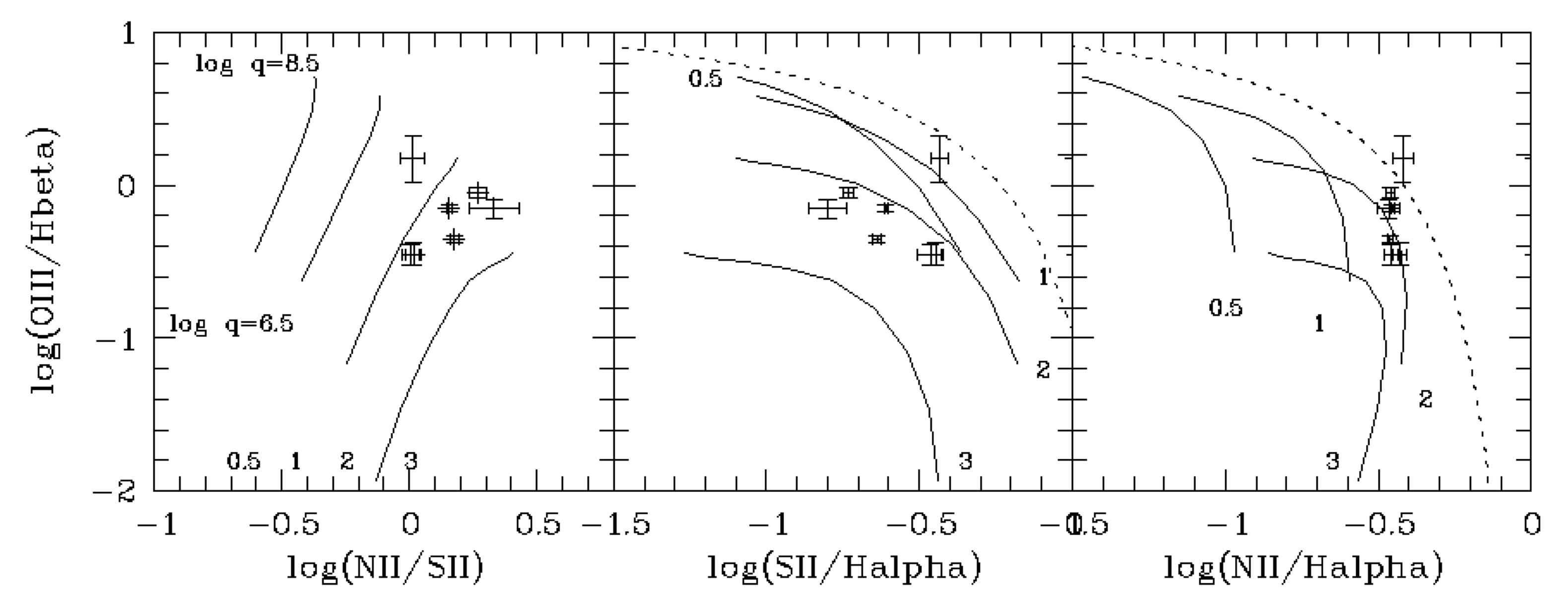} 
\caption{Diagnostic graphs for six selected emission line spectra in SH2 (see Table~\ref{tab:ratios} and Section~\ref{sec:char} for explanation). These spectra are compared to H{\sc ii} region models from \citet{dopita13} with the ionisation parameter $\log q$ and   four oxygen abundances ($0.5\times$ solar, solar, $2\times$ solar, and $3\times$ solar) as parameters. The solid lines are lines of constant oxygen abundance and varying $\log q$. In the left panel, the range of $\log q$ values is indicated, and is the same for all panels. The models refer to $\kappa=\infty$ (corresponding to a Boltzmann electron distribution). The dotted lines in the middle and right panels separate H{\sc ii} regions from AGN- and LINER-spectra according to \citet{kewley01}. All spectra except 25:17 are consistent with a uniform abundance value, although [O{\sc iii}]/H$\beta$ varies significantly. The middle and the right graph show  the degeneracy for high metallicities that is not present in the left graph. A super-solar oxygen abundance is supported by all three diagnostic graphs.}
\label{fig:diagnostics}
\end{center}
\end{figure*}

The precise determination of  element abundances in H{\sc ii} regions normally relies on the determination of electron temperatures, where [O{\sc iii}]-4636\AA\ is the most important line. However, this line is not visible in any of our spectra, which indicates a low temperature and therefore a high metallicity with enhanced cooling capability. This also fits  the fact that neutral He-lines are visible. Consequently, we use strong-line diagnostic diagrams to estimate the oxygen abundance of SH2. There is a huge amount of  literature on H{\sc ii} regions and how to use strong-line diagnostic diagrams. We refer to \cite{dopita13} and \cite{nicholls14} as the primary links with the history of H{\sc ii} region diagnostics. These authors revisit the physics of H{\sc ii} regions, using the most recent atomic line parameters, with the motivation that the energy distribution of electrons might be a $\kappa$-distribution rather than Boltzmann-Maxwellian ($\kappa = \infty$).
These works present many diagnostic diagrams using strong lines, of which only a few are useful for us  due to  the restricted wavelength range.
Furthermore, reliable line-ratios are preferably constructed from lines  from either HRB or HRR. This leaves us with the diagnostic line ratios [O{\sc iii}]/H$\beta$, [N{\sc ii}]/H$\alpha$, [S{\sc ii}]/H$\alpha$, and [N{\sc ii}/S{\sc ii}], which correspond to the diagrams Fig. 9, Fig. 10, and Fig. 22 of \citet{dopita13} (in the case of [N{\sc ii}], we refer to the 6548\AA\ \-line, in the case of [S{\sc ii}] to
the sum of the 6715\AA\ and 6731\AA\ lines). The [Ar{\sc iii}]-line at 7135\AA\ is too weak to permit sensible measurements. The line ratios and the locations of our selected spectra in these diagnostic diagrams are shown in Table~\ref{tab:ratios} and in Fig.~\ref{fig:diagnostics}. Their uncertainties are calculated by error propagations of the 
uncertainties shown in Tables~\ref{tab:fluxesHRB} and \ref{tab:fluxesHRR}. 

\begin{table*}[t]
\centering
\caption{Diagnostic line ratios (Cols. 3 - 7) and results from empirical calibrations (Cols. 8 -  10). See Sect.~\ref{sec:char} for the properties of the spectra. The ID numbers are identified in Fig.~\ref{fig:diagnostics}.}.
\label{tab:ratios}
\begin{tabular}{ccrcrrcccc}
\hline
\hline
ID & spectrum   & log [O{\sc iii}]/H$\beta$ & log [N{\sc ii}/S{\sc ii}] & log [S{\sc ii}]/H$\alpha$ & log [N{\sc ii}]/H$\alpha$\, (N2)& O3N2 & Z(N2) & Z(O3N2) &Z(PG) \\
 \hline
1 & 21:21 &   $-0.15\pm0.03$ & $0.16\pm0.03$ & $-0.61\pm0.01$ & $-0.45\pm0.01$ & $0.30\pm0.03$ & $8.53\pm0.03$  & $8.46\pm0.01$  & 8.57 \\
2 &17:17  &   $-0.05\pm0.03$ & $0.27\pm0.11$ & $-0.73\pm0.03$ & $-0.46\pm0.02$ &$ 0.41\pm0.04$ &  $8.53\pm0.03$ & $8.44\pm0.03$ &  8.58\\
 3 &  4:5   &   $-0.15\pm0.06$ & $0.33\pm0.31$ & $-0.80\pm0.14$ & $-0.47\pm0.08$ &$ 0.31\pm0.10$ & $8.53\pm0.03$ & $8.47\pm0.03$  & 8.60 \\
4 & 25:17 &   $ 0.17\pm0.15$ & $0.01\pm0.13$ & $-0.43\pm0.04$ & $-0.42\pm0.05$ &$ 0.59\pm0.16$ &  $8.55\pm0.03$ & $8.41\pm0.03$ & 8.56 \\
5 & 22:15 &   $-0.45\pm0.07$ & $0.02\pm0.08$ & $-0.45\pm0.03$ & $-0.43\pm0.03$ &$-0.02\pm0.08$ & $8.54\pm0.03$  & $8.53\pm0.03$ &  8.53  \\
6 &20:15  &   $-0.36\pm0.02$ & $0.18\pm0.04$ & $-0.64\pm0.03$ & $-0.46\pm0.02$ &$ 0.11\pm0.03$ &  $8.53\pm0.03$  &  $8.51\pm0.03$ & 8.56  \\  
7 & 21:20  &   $-0.46\pm0.03$ & $0.18\pm0.04$ & $-0.64\pm0.03$ & $-0.46\pm0.02$ &$ 0.11\pm0.03$ &  $ 8.53\pm0.03$  &$8.53\pm0.03$   &  8.60  \\
 \hline\hline
\end{tabular}
\end{table*}

The models are taken from \citet{dopita13} and use the ionisation parameter $\log q$ and the oxygen abundance as parameters. The abundances in solar units are indicated in all panels by the numbers 0.5, 1, 2, 3. The interval of $\log q$ values, which is the same for all models, is indicated as well. According to the definition of \citet{dopita13}, $q$ is the number of ionising photons passing through a unit area divided by the number density of neutral atoms and ions. High $q$-values are therefore realised when the ionising sources are nearby (planetary nebulae) and/or when the spectrum of ionising photons is harder than that produced by an OB stellar population. Another effect is that for high metallicities, some indices are folded back onto the regime of lower metallicities because the enhanced cooling diminishes the fraction of ionised species with respect to hydrogen. Therefore, there is a limit that line ratios of H{\sc ii} regions normally do not cross and which can be used to distinguish the line ratios of H{\sc ii} regions from those of AGNs and LINERs. These limits (which in reality are not very strict) are taken from \citet{kewley01}, and are indicated for the middle and the right panel by the dotted lines. The $\log q$ and abundances are best separated in the left panel, where the lines of constant $\log q$ are almost vertical. It is satisfactory that although $\log q$ varies, the loci of the spectrum follow a constant abundance line with the exception of 25-17. This is also the case in the middle panel, although there is the above-mentioned degeneracy, in the sense that metal-rich loci cross with the metal-poor ones. The right panel indicates a somewhat lower abundance. 

For the interpolation in the tables of \citet{dopita13}, we use linear expressions in the intervals of interest, which are log q $>$ 7.0 and  1$<$z$\le$3.    For the left panel, we obtain (valid only for this selection)
\begin{equation}
\begin{split}
z = 1.8 (\pm0.06) +1.8(\pm0.1) \times log [NII/SII] \\
 -1.1(\pm0.03) \times log [OIII]
\end{split}
\end{equation}
For the middle panel, we obtain
\begin{equation}
\begin{split}
z = 1.44(\pm0.06) -1.42(\pm0.1) \times log [SII/H\alpha] \\
 -1.4(\pm0.08) \times log [OIII]
\end{split}
\end{equation}
For the right panel, we adopt the value z=2.0 with the exception of spectrum 4. 

Neglecting spectrum 4, the resulting  oxygen abundances (in solar units) are 2.44$\pm$0.15  for the left panel  and 2.31$\pm$0.2 for  the middle panel.  With  the
the solar abundance  12 + log(O/H)=8.69 \citep{grevesse10}, these values transform  to   12 + log(O/H)=9.08 and 9.05, respectively. The right panel gives 12 + log(O/H)=8.99.
 
A comparison with the parameters of the H{\sc ii} regions of \citet{vanzee98}, provided by \citeauthor{dopita13} in their graphs, shows that SH2 belongs to the H{\sc ii} regions with the highest abundance and highest ionisation parameters known.  Spectrum 4, the spectrum with the highest ionisation parameter, deviates in that the abundance is not consistent with the other spectra. As mentioned before, the position is consistent with PN 397 in the catalogue of \citet{mcneil12}. The velocity from this
catalogue is 1731$\pm$30 km/s, also consistent with an affiliation to SH2. However, there are more PNe in this region with similar velocities that certainly do not
belong to SH2. Moreover, the  [O{\sc iii}]/H$\beta$ ratio is quite small compared with other planetary nebulae.  We cannot decide whether this is a true planetary nebula
or a particularly high ionisation flux in the HII region, and we leave it as an interesting case.


\subsection{Empirical calibrations of oxygen abundance}

To date  there is no commonly accepted metallicity scale for H{\sc ii} regions. Abundances that have been obtained via determinations of electron temperatures
are often lower by 0.3 - 0.5 dex than  those that have been obtained via photoionisation models (e.g. \citealt{bresolin09}; see also the discussion section in \citealt{pilyugin16}). The reasons are still under debate. Indeed, we also see this discrepancy  in our data when we apply empirical calibrations.

\subsubsection{Calibration by Marino et al. 2013}
\citet{marino13} use the CALIFA data set and temperature-based abundances in the literature, to update the calibrations for the indicators N2 and O3N2, being 
\begin{eqnarray*}
{\rm N2}     &=& \log({\rm [N{\sc II}]6584/H}\alpha) \,\,\, {\rm and} \\
{\rm O3N2}&=& \log\left\{({\rm [O{III}]5007/H}\beta)/([{\rm N{II}]6584/H}_\alpha)\right\}.
\end{eqnarray*}
In these calibrations, H{\sc ii} regions appear with the abundance as the only parameter, which in the case of high metallicity is degenerated. The proposed calibrations are

\begin{equation}
\label{eq:O3N2}
12+\log({\rm O/H}) = 8.533 - 0.214\times {\rm O3N2}
\end{equation}

\noindent and

\begin{equation}
\label{eq:N2}
12+\log({\rm O/H}) = 8.743 + 0.462 \times {\rm N2}
\end{equation}

The values of these parameters for our spectra are listed in Cols.  6 - 9 in Table~\ref{tab:ratios}. The  calibrations result in mean oxygen abundances of 12+log[O/H] = $8.53\pm0.009$ for N2 and 12+log[O/H] = $8.48\pm0.05$ for O3N2, 
 where the uncertainties have been calculated by error propagation using the uncertainties in the calibrating relations. As expected, these values are systematically lower, by approximately 0.5 dex, than those obtained in Sect. 6.1 from photoionisation models. For the following discussion, however,  it is  worth pointing out that even the empirical calibrations indicate an oxygen abundance corresponding to about 60\% solar.

\subsubsection{ Calibration of Pilyugin \& Grebel 2016}
An empirical calibration  with a small scatter of 0.1 dex has been presented by \citet{pilyugin16}, who use temperature-based abundances of H{\sc ii} regions to construct relations between three strong-line indices and abundances. They calibrate the following indices in terms of oxygen abundance: N2 = $(6548+6583)/$H$\beta$, S2 = $(6715+6731)/$H$\beta$, R3 = $(4958+5007)/$H$\beta$, where the wavelengths stand for the fluxes of the respective lines (they calibrate additional indices involving [O{\sc ii}] 3727\AA\ that are not relevant for us).  Only for R3  can we measure all fluxes using the same spectrum, and so we adopt H$\alpha\!=\!2.89\times$H$\beta$ for N2 and S2. We use equation (6) of \citet{pilyugin16} to derive for our spectra a mean oxygen abundance of 12+log[O/H] = $8.55\pm0.05$, where the uncertainty is the standard deviation. 

The abundances from the calibrations using T$_e$-based methods are thus in very good mutual agreement, but  distinctly lower
than the values from the diagnostic graphs. However,  irrespective of the absolute value, an abundance of 12+log[O/H] = 8.55 still belongs to the most metal-rich HII regions
in the calibrating sample of \citet{pilyugin16} where the extreme value is 8.7.

  
\section{Discussion}


\subsection{Possible formation and evolution of SH2}
\label{sec:scenario}
The high oxygen abundance of SH2 which is indicated by both the line diagnostics and empirical calibrations, excludes the nature of SH2 as a dwarf galaxy. 
According to the mass-metallicity relation for dwarf galaxies,    an oxygen abundance of 12+log(O/H) = 7.74 (solar abundance 12+log(O/H) = 8.69) is expected for a stellar mass of $10^6 M_\odot$, or 12+log(O/H) = 7.85 if the total H{\sc i} mass of $10^7 M_\odot$ is inserted \citep[e.g.][]{izotov15}, aside from the fact that such an extremely gas-rich dwarf galaxy would be quite exotic, in particular in close vicinity to a giant  galaxy such as NGC\,1316. This high abundance of SH2 rather fits to solar metallicities that have been determined for some bright globular clusters in NGC\,1316 \citep{goudfrooij01b}. These clusters were probably   formed  during a period of very high star formation  $\sim\!2$ Gyr ago which generated a major part of the field stellar population \citep{richtler12a}. However, globular cluster formation did not cease after the main starburst.  We found one globular cluster (there are perhaps more) with an age of about 0.5 Gyr (the object n1316-gc01178 of \citealt{richtler12a}) without an obvious field component of this age \citep{richtler14}. The complex SH2, after being dispersed, will provide the globular cluster system of NGC\,1316 with more young star clusters without contributing much to the field stellar population.

Since there is no evidence for recent star formation in NGC\,1316 and no direct evidence for the existence of larger molecular cloud complexes outside the central regions \citep{lanz10}, the parent cloud of SH2 must have been quite isolated, but was not unique in NGC\,1316. A few arcminutes to the north, \citet{mackie98} found an ``extended emission line region (EELR)" . 
The ionising photons seemingly do not come from a young stellar population as in the case of SH2. The H{\sc i} mass is about $10^7 M_\odot$, similar to SH2. The colour map shown by \citet{richtler12b} reveals a reddish patch, indicative of dust. 
The basic difference between SH2 and the EELR seems to be the occurrence of star formation. It can be that the EELR is in an earlier evolutionary stage than SH2 and a starburst is still to come. 
 
The question is whether it needs an external trigger like ram pressure or a tidal shock to commence star formation. We find a similar situation in many blue compact dwarfs, which show recent clustered star formation even in very isolated locations after a long period of quiescence \citep{fuse12}. 
We also mention the possibility of ``jet-induced star formation'' that has been discussed, for example  to explain the existence of a young stellar population near Centaurus A at
a galactocentric distance of 15 kpc \citep{mould00,santoro16,salome16}.  In that case the age of SH2 would indicate the epoch of the last nuclear activity in NGC1316.
 The fact that the radio jet in NGC1316 and the line nucleus-SH2 are strongly misaligned is perhaps no counterargument  because of the differential rotation and precession
 of the jet. However, there is no further evidence for this interesting process. 

A possible scenario is that of a slow long-term evolution of molecular material where the critical density for star formation evolves over a time span of several Gyrs. That a molecular cloud complex survives without star formation much longer than about $10^8$ years is not easily understandable, so it is more probable that molecules formed inside the densest parts of an original H{\sc i} complex during a long period of $1\!-\!2$ Gyr. Once a molecular cloud core has formed and is shielded, there is hardly anything that can prevent star formation from beginning \citep[e.g.][]{elmegreen07} and many bound star clusters can form in high-density peaks \citep[see e.g.][]{kruijssen12}.

To understand the nature of SH2, it would be helpful to compare it with other objects. But regarding the combination of the principal characteristics such as extreme clustering, ring-like morphology, and isolation, SH2 seems to be unique in the literature. This would not be surprising, if this kind of star formation were exclusively  found in the aftermath of mergers. To our knowledge, even nearby prominent mergers like NGC\,474 are not surveyed to this detail. 
However, considering these properties separately,   extreme clustering is also found in the starbursts of dwarf galaxies \citep{adamo11}, for example in Haro\,11 \citep{adamo10} where even more clusters than in SH2 are crowded together within a similar volume. A ring-like cluster configuration, known as the Ruby Ring, has been identified in NGC\,2146 \citep{adamo12}. The authors offer an explanation based on the existence of a central massive star cluster where supernovae and stellar winds sweep up the interstellar medium to trigger secondary star formation in a ring. In the case of SH2, an obvious central object is missing, but one may speculate that $10^8$ years ago, the brightest object 431, now displaced, was the central cluster. In that case, one would expect the ``ring population'' to be considerably younger, which at least for the ionising population apparently is the case.

\subsection{A failed UCD?}

Intermediate-age star clusters are found in NGC1316 with  masses as high as about $2\times10^7 M_\odot$ \citep{goudfrooij01b,richtler14} that accordingly can be labeled Ultracompact Dwarfs (UCDs) (e.g. \citealt{hilker07}).  
The global star formation efficiency in SH2 is quite low considering the total gas mass of about $10^7 M_\odot$. We do not know why a more massive object
has not  formed, but we can estimate that the present configuration is unlikely to merge and form one object. The ring-like appearance strongly indicates a gas disk at the
time of formation. If we assume as an extreme that the total mass is inside the ring, we expect a rotation velocity of about 10 km/s, probably lower than the velocity
dispersion within this zone of NGC1316. We therefore expect it to be a transient configuration that will disperse.

\subsection{ Role of the star cluster complex 429}
\label{sec:429}

The intermediate-age massive globular cluster 429 is not easily brought into connection with SH2. If the three-dimensional structure in the region, including globular clusters, was more or less isotropic, the agreement of the radial velocities of this object and SH2, including all line emission, would be a very surprising coincidence. If, on the other hand, we consider  a large-scale disk,
the difficulty of explaining the connection of 429 with SH2 disappears.
 In this case, 429 would only  be  a cluster casually orbiting nearby.   A  hint might be the peak in the  distribution of globular cluster velocities at about 1600 km/s, which is in
 quite good agreement with the velocity  of 429 \citep{richtler12b} and may be interpreted as a disk population. A stronger constraint can come from the kinematics of the galaxy light
 itself, because a  disk  should be recognisable by its low velocity dispersion. However, our signal-to-noise ratio is too low for this kind of analysis.
 
 Another interesting feature is that this 2 Gyr old cluster is apparently capable of ionising its environment which  may  seem surprising at first. However, 
post-AGB stars  of much older populations ionise gas in early-type galaxies \citep{stasinska08,cid11,kehrig12,johansson14}. According to Fig. 2 of \citet{cid11}, the flux of HI-ionising photons increases by two orders of magnitudes at an age of about 0.1 Gyr and then remains at an almost constant level.  
 We are not aware of any other example in the literature that indicates an ionising photons flux from an intermediate-age globular cluster.

\section{Summary and conclusions}
We have investigated the emission lines of the isolated H{\sc ii} region and star cluster complex SH2 in the merger remnant NGC\,1316 with data from the VIMOS/IFU at the Very Large Telescope. Our objective was to test the hypothesis of SH2 being an infalling dwarf galaxy. We used strong-line diagnostic diagrams and empirical calibrations to determine the metallicity of the gas and arrive at 
a nearly solar or even super-solar oxygen abundance. 
The high metallicity of SH2 disfavours the interpretation of an infalling dwarf galaxy and favours a scenario where a molecular cloud complex, perhaps formed during the merger $\sim\!2$ Gyr ago, was quiescent for a long time and started to form stars about 0.1 Gyr ago. 
Its isolated character might come from a dynamically chaotic situation during a merger/infall event when large coherent structures (perhaps molecular in nature)  were disrupted and started their own dynamical life. A long-term evolution of the molecular material then resulted finally in a critical density for star formation perhaps without an external triggering event. 
This also explains the existence of a few massive star clusters in NGC\,1316 with ages of about 0.5 Gyr without a significant field component.  

The association with the intermediate-age cluster 429 remains unclear. The concordant radial velocities might reflect a disk-like kinematic seen face-on. The dense clustering is also found in star-forming dwarf galaxies, and even the ring-like morphology finds its analogue in NGC\,2146. SH2 is an illustrative example of massive star  cluster formation outside of starburst periods. 

\begin{acknowledgements}
We thank the referee, S{\o}ren S. Larsen, for a constructive report and helpful suggestions that improved the paper.
T.R. acknowledges support 
from  the BASAL Centro de
 Astrof\'{\i}sica y Tecnologias Afines (CATA) PFB-06/2007.  T.R. thanks   ESO/Garching for a  science visitorship that was essential for the present work. ~ T.H.P.\ acknowledges support by a FONDECYT Regular Project Grant (No.~1161817) and the BASAL Center for Astrophysics and Associated Technologies (PFB-06).
\end{acknowledgements}

\appendix

\section{Spectra}

\begin{figure*}
\begin{center}
\includegraphics[width=0.9\textwidth]{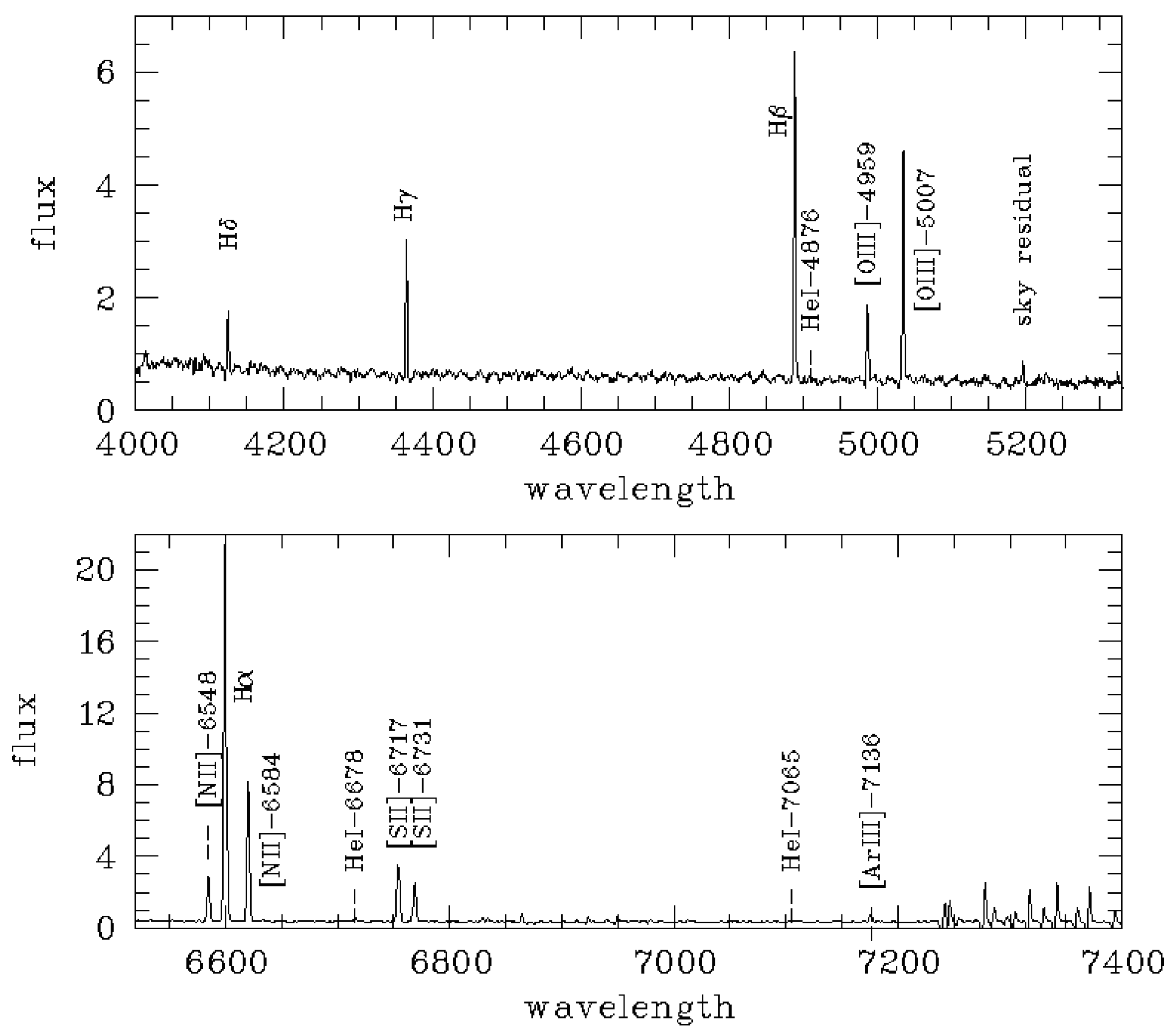}
\caption{ Total spectrum. The flux is the sum of an extraction centred on 21:21 and summed up within a radius of eight pixels. 
The upper panel shows the spectrum from the grism HRB, the lower from the grism HRR.
The visibility of  He{\sc i} lines and the absence of the O{\sc ii}-4636 line already indicate a low temperature and accordingly a high metallicity, which is
confirmed by a more detailed line diagnostics.  }
\label{fig:spectrum_total}
\end{center}
\end{figure*}





\bibliographystyle{aa}
\bibliography{paper_V18_astroph.bib}

\begin{thebibliography}{55}
\expandafter\ifx\csname natexlab\endcsname\relax\def\natexlab#1{#1}\fi

\bibitem[{{Adamo} {et~al.}(2011){Adamo}, {{\"O}stlin}, \&
  {Zackrisson}}]{adamo11}
{Adamo}, A., {{\"O}stlin}, G., \& {Zackrisson}, E. 2011, \mnras, 417, 1904

\bibitem[{{Adamo} {et~al.}(2010){Adamo}, {{\"O}stlin}, {Zackrisson}, {Hayes},
  {Cumming}, \& {Micheva}}]{adamo10}
{Adamo}, A., {{\"O}stlin}, G., {Zackrisson}, E., {et~al.} 2010, \mnras, 407,
  870

\bibitem[{{Adamo} {et~al.}(2012){Adamo}, {Smith}, {Gallagher}, {Bastian},
  {Ryon}, {Westmoquette}, {Konstantopoulos}, {Zackrisson}, {Larsen},
  {Silva-Villa}, {Charlton}, \& {Weisz}}]{adamo12}
{Adamo}, A., {Smith}, L.~J., {Gallagher}, J.~S., {et~al.} 2012, \mnras, 426,
  1185

\bibitem[{{Arnaboldi} {et~al.}(1998){Arnaboldi}, {Freeman}, {Gerhard},
  {Matthias}, {Kudritzki}, {M{\'e}ndez}, {Capaccioli}, \& {Ford}}]{arnaboldi98}
{Arnaboldi}, M., {Freeman}, K.~C., {Gerhard}, O., {et~al.} 1998, \apj, 507, 759

\bibitem[{{Bedregal} {et~al.}(2006){Bedregal}, {Arag{\'o}n-Salamanca},
  {Merrifield}, \& {Milvang-Jensen}}]{bedregal06}
{Bedregal}, A.~G., {Arag{\'o}n-Salamanca}, A., {Merrifield}, M.~R., \&
  {Milvang-Jensen}, B. 2006, \mnras, 371, 1912

\bibitem[{{Bresolin} {et~al.}(2009){Bresolin}, {Gieren}, {Kudritzki},
  {Pietrzy{\'n}ski}, {Urbaneja}, \& {Carraro}}]{bresolin09}
{Bresolin}, F., {Gieren}, W., {Kudritzki}, R.-P., {et~al.} 2009, \apj, 700, 309

\bibitem[{{Cantiello} {et~al.}(2013){Cantiello}, {Grado}, {Blakeslee},
  {Raimondo}, {Di Rico}, {Limatola}, {Brocato}, {Della Valle}, \&
  {Gilmozzi}}]{cantiello13}
{Cantiello}, M., {Grado}, A., {Blakeslee}, J.~P., {et~al.} 2013, \aap, 552,
  A106

\bibitem[{{Cid Fernandes} {et~al.}(2011){Cid Fernandes}, {Stasi{\'n}ska},
  {Mateus}, \& {Vale Asari}}]{cid11}
{Cid Fernandes}, R., {Stasi{\'n}ska}, G., {Mateus}, A., \& {Vale Asari}, N.
  2011, \mnras, 413, 1687

\bibitem[{{D'Onofrio} {et~al.}(1995){D'Onofrio}, {Zaggia}, {Longo}, {Caon}, \&
  {Capaccioli}}]{donofrio95}
{D'Onofrio}, M., {Zaggia}, S.~R., {Longo}, G., {Caon}, N., \& {Capaccioli}, M.
  1995, \aap, 296, 319

\bibitem[{{Dopita} {et~al.}(2013){Dopita}, {Sutherland}, {Nicholls}, {Kewley},
  \& {Vogt}}]{dopita13}
{Dopita}, M.~A., {Sutherland}, R.~S., {Nicholls}, D.~C., {Kewley}, L.~J., \&
  {Vogt}, F.~P.~A. 2013, \apjs, 208, 10

\bibitem[{{Elmegreen}(2007)}]{elmegreen07}
{Elmegreen}, B.~G. 2007, \apj, 668, 1064

\bibitem[{{Fuse} {et~al.}(2012){Fuse}, {Marcum}, \& {Fanelli}}]{fuse12}
{Fuse}, C., {Marcum}, P., \& {Fanelli}, M. 2012, \aj, 144, 57

\bibitem[{{G{\'o}mez} {et~al.}(2001){G{\'o}mez}, {Richtler}, {Infante}, \&
  {Drenkhahn}}]{gomez01}
{G{\'o}mez}, M., {Richtler}, T., {Infante}, L., \& {Drenkhahn}, G. 2001, \aap,
  371, 875

\bibitem[{{Goudfrooij}(2012)}]{goudfrooij12}
{Goudfrooij}, P. 2012, \apj, 750, 140

\bibitem[{{Goudfrooij} {et~al.}(2001{\natexlab{a}}){Goudfrooij}, {Alonso},
  {Maraston}, \& {Minniti}}]{goudfrooij01a}
{Goudfrooij}, P., {Alonso}, M.~V., {Maraston}, C., \& {Minniti}, D.
  2001{\natexlab{a}}, \mnras, 328, 237

\bibitem[{{Goudfrooij} {et~al.}(2001{\natexlab{b}}){Goudfrooij}, {Mack},
  {Kissler-Patig}, {Meylan}, \& {Minniti}}]{goudfrooij01b}
{Goudfrooij}, P., {Mack}, J., {Kissler-Patig}, M., {Meylan}, G., \& {Minniti},
  D. 2001{\natexlab{b}}, \mnras, 322, 643

\bibitem[{{Grevesse} {et~al.}(2010){Grevesse}, {Asplund}, {Sauval}, \&
  {Scott}}]{grevesse10}
{Grevesse}, N., {Asplund}, M., {Sauval}, A.~J., \& {Scott}, P. 2010, \apss,
  328, 179

\bibitem[{{Hilker} {et~al.}(2007){Hilker}, {Baumgardt}, {Infante},
  {Drinkwater}, {Evstigneeva}, \& {Gregg}}]{hilker07}
{Hilker}, M., {Baumgardt}, H., {Infante}, L., {et~al.} 2007, \aap, 463, 119

\bibitem[{{Horellou} {et~al.}(2001){Horellou}, {Black}, {van Gorkom}, {Combes},
  {van der Hulst}, \& {Charmandaris}}]{horellou01}
{Horellou}, C., {Black}, J.~H., {van Gorkom}, J.~H., {et~al.} 2001, \aap, 376,
  837

\bibitem[{{Husemann} {et~al.}(2016){Husemann}, {Bennert}, {Scharw{\"a}chter},
  {Woo}, \& {Choudhury}}]{husemann16}
{Husemann}, B., {Bennert}, V.~N., {Scharw{\"a}chter}, J., {Woo}, J.-H., \&
  {Choudhury}, O.~S. 2016, \mnras, 455, 1905

\bibitem[{{Husemann} {et~al.}(2013){Husemann}, {Jahnke}, {S{\'a}nchez},
  {Barrado}, {Bekerait*error*{\.e}}, {Bomans}, {Castillo-Morales},
  {Catal{\'a}n-Torrecilla}, {Cid Fernandes}, {Falc{\'o}n-Barroso},
  {Garc{\'{\i}}a-Benito}, {Gonz{\'a}lez Delgado}, {Iglesias-P{\'a}ramo},
  {Johnson}, {Kupko}, {L{\'o}pez-Fernandez}, {Lyubenova}, {Marino}, {Mast},
  {Miskolczi}, {Monreal-Ibero}, {Gil de Paz}, {P{\'e}rez}, {P{\'e}rez},
  {Rosales-Ortega}, {Ruiz-Lara}, {Schilling}, {van de Ven}, {Walcher}, {Alves},
  {de Amorim}, {Backsmann}, {Barrera-Ballesteros}, {Bland-Hawthorn}, {Cortijo},
  {Dettmar}, {Demleitner}, {D{\'{\i}}az}, {Enke}, {Florido}, {Flores},
  {Galbany}, {Gallazzi}, {Garc{\'{\i}}a-Lorenzo}, {Gomes}, {Gruel}, {Haines},
  {Holmes}, {Jungwiert}, {Kalinova}, {Kehrig}, {Kennicutt}, {Klar}, {Lehnert},
  {L{\'o}pez-S{\'a}nchez}, {de Lorenzo-C{\'a}ceres}, {M{\'a}rmol-Queralt{\'o}},
  {M{\'a}rquez}, {Mendez-Abreu}, {Moll{\'a}}, {del Olmo}, {Meidt}, {Papaderos},
  {Puschnig}, {Quirrenbach}, {Roth}, {S{\'a}nchez-Bl{\'a}zquez}, {Spekkens},
  {Singh}, {Stanishev}, {Trager}, {Vilchez}, {Wild}, {Wisotzki}, {Zibetti}, \&
  {Ziegler}}]{husemann13}
{Husemann}, B., {Jahnke}, K., {S{\'a}nchez}, S.~F., {et~al.} 2013, \aap, 549,
  A87

\bibitem[{{Husemann} {et~al.}(2014){Husemann}, {Jahnke}, {S{\'a}nchez},
  {Wisotzki}, {Nugroho}, {Kupko}, \& {Schramm}}]{husemann14}
{Husemann}, B., {Jahnke}, K., {S{\'a}nchez}, S.~F., {et~al.} 2014, \mnras, 443,
  755

\bibitem[{{Husemann} {et~al.}(2012){Husemann}, {Kamann}, {Sandin},
  {S{\'a}nchez}, {Garc{\'{\i}}a-Benito}, \& {Mast}}]{husemann12}
{Husemann}, B., {Kamann}, S., {Sandin}, C., {et~al.} 2012, \aap, 545, A137

\bibitem[{{Izotov} {et~al.}(2015){Izotov}, {Guseva}, {Fricke}, \&
  {Henkel}}]{izotov15}
{Izotov}, Y.~I., {Guseva}, N.~G., {Fricke}, K.~J., \& {Henkel}, C. 2015,
  \mnras, 451, 2251

\bibitem[{{Johansson} {et~al.}(2014){Johansson}, {Woods}, {Gilfanov}, {Sarzi},
  {Chen}, \& {Oh}}]{johansson14}
{Johansson}, J., {Woods}, T.~E., {Gilfanov}, M., {et~al.} 2014, \mnras, 442,
  1079

\bibitem[{{Kehrig} {et~al.}(2012){Kehrig}, {Monreal-Ibero}, {Papaderos},
  {V{\'{\i}}lchez}, {Gomes}, {Masegosa}, {S{\'a}nchez}, {Lehnert}, {Cid
  Fernandes}, {Bland-Hawthorn}, {Bomans}, {Marquez}, {Mast}, {Aguerri},
  {L{\'o}pez-S{\'a}nchez}, {Marino}, {Pasquali}, {Perez}, {Roth},
  {S{\'a}nchez-Bl{\'a}zquez}, \& {Ziegler}}]{kehrig12}
{Kehrig}, C., {Monreal-Ibero}, A., {Papaderos}, P., {et~al.} 2012, \aap, 540,
  A11

\bibitem[{{Kewley} {et~al.}(2001){Kewley}, {Dopita}, {Sutherland}, {Heisler},
  \& {Trevena}}]{kewley01}
{Kewley}, L.~J., {Dopita}, M.~A., {Sutherland}, R.~S., {Heisler}, C.~A., \&
  {Trevena}, J. 2001, \apj, 556, 121

\bibitem[{{Kim} \& {Fabbiano}(2003)}]{kim03}
{Kim}, D.-W. \& {Fabbiano}, G. 2003, \apj, 586, 826

\bibitem[{{Kruijssen}(2012)}]{kruijssen12}
{Kruijssen}, J.~M.~D. 2012, \mnras, 426, 3008

\bibitem[{{Kuntschner}(2000)}]{kuntschner00}
{Kuntschner}, H. 2000, \mnras, 315, 184

\bibitem[{{Lanz} {et~al.}(2010){Lanz}, {Jones}, {Forman}, {Ashby}, {Kraft}, \&
  {Hickox}}]{lanz10}
{Lanz}, L., {Jones}, C., {Forman}, W.~R., {et~al.} 2010, \apj, 721, 1702

\bibitem[{{Le F{\`e}vre} {et~al.}(2003){Le F{\`e}vre}, {Saisse}, {Mancini},
  {Brau-Nogue}, {Caputi}, {Castinel}, {D'Odorico}, {Garilli}, {Kissler-Patig},
  {Lucuix}, {Mancini}, {Pauget}, {Sciarretta}, {Scodeggio}, {Tresse}, \&
  {Vettolani}}]{lefevre03}
{Le F{\`e}vre}, O., {Saisse}, M., {Mancini}, D., {et~al.} 2003, in \procspie,
  Vol. 4841, Instrument Design and Performance for Optical/Infrared
  Ground-based Telescopes, ed. M.~{Iye} \& A.~F.~M. {Moorwood}, 1670--1681

\bibitem[{{Longhetti} {et~al.}(1998){Longhetti}, {Rampazzo}, {Bressan}, \&
  {Chiosi}}]{longhetti98}
{Longhetti}, M., {Rampazzo}, R., {Bressan}, A., \& {Chiosi}, C. 1998, \aaps,
  130, 267

\bibitem[{{Mackie} \& {Fabbiano}(1998)}]{mackie98}
{Mackie}, G. \& {Fabbiano}, G. 1998, \aj, 115, 514

\bibitem[{{Marino} {et~al.}(2013){Marino}, {Rosales-Ortega}, {S{\'a}nchez},
  {Gil de Paz}, {V{\'{\i}}lchez}, {Miralles-Caballero}, {Kehrig},
  {P{\'e}rez-Montero}, {Stanishev}, {Iglesias-P{\'a}ramo}, {D{\'{\i}}az},
  {Castillo-Morales}, {Kennicutt}, {L{\'o}pez-S{\'a}nchez}, {Galbany},
  {Garc{\'{\i}}a-Benito}, {Mast}, {Mendez-Abreu}, {Monreal-Ibero}, {Husemann},
  {Walcher}, {Garc{\'{\i}}a-Lorenzo}, {Masegosa}, {Del Olmo Orozco},
  {Mour{\~a}o}, {Ziegler}, {Moll{\'a}}, {Papaderos},
  {S{\'a}nchez-Bl{\'a}zquez}, {Gonz{\'a}lez Delgado}, {Falc{\'o}n-Barroso},
  {Roth}, {van de Ven}, \& {Califa Team}}]{marino13}
{Marino}, R.~A., {Rosales-Ortega}, F.~F., {S{\'a}nchez}, S.~F., {et~al.} 2013,
  \aap, 559, A114

\bibitem[{{McNeil-Moylan} {et~al.}(2012){McNeil-Moylan}, {Freeman},
  {Arnaboldi}, \& {Gerhard}}]{mcneil12}
{McNeil-Moylan}, E.~K., {Freeman}, K.~C., {Arnaboldi}, M., \& {Gerhard}, O.~E.
  2012, ArXiv e-prints

\bibitem[{{Mould} {et~al.}(2000){Mould}, {Ridgewell}, {Gallagher}, {Bessell},
  {Keller}, {Calzetti}, {Clarke}, {Trauger}, {Grillmair}, {Ballester},
  {Burrows}, {Krist}, {Crisp}, {Evans}, {Griffiths}, {Hester}, {Hoessel},
  {Holtzman}, {Scowen}, {Stapelfeldt}, {Sahai}, {Watson}, \&
  {Meadows}}]{mould00}
{Mould}, J.~R., {Ridgewell}, A., {Gallagher}, III, J.~S., {et~al.} 2000, \apj,
  536, 266

\bibitem[{{Nicholls} {et~al.}(2014){Nicholls}, {Dopita}, {Sutherland},
  {Jerjen}, {Kewley}, \& {Basurah}}]{nicholls14}
{Nicholls}, D.~C., {Dopita}, M.~A., {Sutherland}, R.~S., {et~al.} 2014, \apj,
  786, 155

\bibitem[{{Nowak} {et~al.}(2008){Nowak}, {Saglia}, {Thomas}, {Bender},
  {Davies}, \& {Gebhardt}}]{nowak08}
{Nowak}, N., {Saglia}, R.~P., {Thomas}, J., {et~al.} 2008, \mnras, 391, 1629

\bibitem[{{Panuzzo} {et~al.}(2003){Panuzzo}, {Bressan}, {Granato}, {Silva}, \&
  {Danese}}]{panuzzo03}
{Panuzzo}, P., {Bressan}, A., {Granato}, G.~L., {Silva}, L., \& {Danese}, L.
  2003, \aap, 409, 99

\bibitem[{{Pilyugin} \& {Grebel}(2016)}]{pilyugin16}
{Pilyugin}, L.~S. \& {Grebel}, E.~K. 2016, \mnras, 457, 3678

\bibitem[{{Richtler} {et~al.}(2012{\natexlab{a}}){Richtler}, {Bassino},
  {Dirsch}, \& {Kumar}}]{richtler12a}
{Richtler}, T., {Bassino}, L.~P., {Dirsch}, B., \& {Kumar}, B.
  2012{\natexlab{a}}, \aap, accepted, Paper I

\bibitem[{{Richtler} {et~al.}(2014){Richtler}, {Hilker}, {Kumar}, {Bassino},
  {G{\'o}mez}, \& {Dirsch}}]{richtler14}
{Richtler}, T., {Hilker}, M., {Kumar}, B., {et~al.} 2014, \aap, 569, A41

\bibitem[{{Richtler} {et~al.}(2012{\natexlab{b}}){Richtler}, {Kumar},
  {Bassino}, {Dirsch}, \& {Romanowsky}}]{richtler12b}
{Richtler}, T., {Kumar}, B., {Bassino}, L.~P., {Dirsch}, B., \& {Romanowsky},
  A.~J. 2012{\natexlab{b}}, \aap, 543, L7

\bibitem[{{Salom{\'e}} {et~al.}(2016){Salom{\'e}}, {Salom{\'e}}, {Combes},
  {Hamer}, \& {Heywood}}]{salome16}
{Salom{\'e}}, Q., {Salom{\'e}}, P., {Combes}, F., {Hamer}, S., \& {Heywood}, I.
  2016, \aap, 586, A45

\bibitem[{{Santoro} {et~al.}(2016){Santoro}, {Oonk}, {Morganti}, {Oosterloo},
  \& {Tadhunter}}]{santoro16}
{Santoro}, F., {Oonk}, J.~B.~R., {Morganti}, R., {Oosterloo}, T.~A., \&
  {Tadhunter}, C. 2016, \aap, 590, A37

\bibitem[{{Schweizer}(1980)}]{schweizer80}
{Schweizer}, F. 1980, \apj, 237, 303

\bibitem[{{Sesto} {et~al.}(2016){Sesto}, {Faifer}, \& {Forte}}]{sesto16}
{Sesto}, L.~A., {Faifer}, F.~R., \& {Forte}, J.~C. 2016, \mnras, 461, 4260

\bibitem[{{Shaya} {et~al.}(1996){Shaya}, {Dowling}, {Currie}, {Faber}, {Ajhar},
  {Lauer}, {Groth}, {Grillmair}, {Lynd}, \& {O'Neil}}]{shaya96}
{Shaya}, E.~J., {Dowling}, D.~M., {Currie}, D.~G., {et~al.} 1996, \aj, 111,
  2212

\bibitem[{{Smith Castelli} {et~al.}(2012){Smith Castelli}, {Cellone}, {Faifer},
  {Bassino}, {Richtler}, {Romero}, {Calder{\'o}n}, \& {Caso}}]{smith12}
{Smith Castelli}, A.~V., {Cellone}, S.~A., {Faifer}, F.~R., {et~al.} 2012,
  \mnras, 419, 2472

\bibitem[{{Stasi{\'n}ska} {et~al.}(2008){Stasi{\'n}ska}, {Vale Asari}, {Cid
  Fernandes}, {Gomes}, {Schlickmann}, {Mateus}, {Schoenell}, {Sodr{\'e}}, \&
  {Seagal Collaboration}}]{stasinska08}
{Stasi{\'n}ska}, G., {Vale Asari}, N., {Cid Fernandes}, R., {et~al.} 2008,
  \mnras, 391, L29

\bibitem[{{Stritzinger} {et~al.}(2010){Stritzinger}, {Burns}, {Phillips},
  {Folatelli}, {Krisciunas}, {Kattner}, {Persson}, {Boldt}, {Campillay},
  {Contreras}, {Krzeminski}, {Morrell}, {Salgado}, {Freedman}, {Hamuy},
  {Madore}, {Roth}, \& {Suntzeff}}]{stritzinger10}
{Stritzinger}, M., {Burns}, C.~R., {Phillips}, M.~M., {et~al.} 2010, \aj, 140,
  2036

\bibitem[{{van Zee} {et~al.}(1998){van Zee}, {Salzer}, {Haynes}, {O'Donoghue},
  \& {Balonek}}]{vanzee98}
{van Zee}, L., {Salzer}, J.~J., {Haynes}, M.~P., {O'Donoghue}, A.~A., \&
  {Balonek}, T.~J. 1998, \aj, 116, 2805

\bibitem[{{Walcher} {et~al.}(2015){Walcher}, {Coelho}, {Gallazzi}, {Bruzual},
  {Charlot}, \& {Chiappini}}]{walcher15}
{Walcher}, C.~J., {Coelho}, P.~R.~T., {Gallazzi}, A., {et~al.} 2015, \aap, 582,
  A46

\bibitem[{{Wuyts} {et~al.}(2011){Wuyts}, {F{\"o}rster Schreiber}, {van der
  Wel}, {Magnelli}, {Guo}, {Genzel}, {Lutz}, {Aussel}, {Barro}, {Berta},
  {Cava}, {Graci{\'a}-Carpio}, {Hathi}, {Huang}, {Kocevski}, {Koekemoer},
  {Lee}, {Le Floc'h}, {McGrath}, {Nordon}, {Popesso}, {Pozzi}, {Riguccini},
  {Rodighiero}, {Saintonge}, \& {Tacconi}}]{wuyts11}
{Wuyts}, S., {F{\"o}rster Schreiber}, N.~M., {van der Wel}, A., {et~al.} 2011,
  \apj, 742, 96

\end{thebibliography}

\end{document}